\documentclass[preprint,5p,times,twocolumn,authoryear]{elsarticle}

\usepackage{amssymb}
\usepackage[dvipsnames]{xcolor} 

\journal{Astronomy and Computing}

\begin{document}

\begin{frontmatter}



\title{Towards Asteroid Detection in Microlensing Surveys with Deep Learning}


\author[inst1]{Preeti Cowan}

\affiliation[inst1]{organization={School of Mathematical and Computational Sciences, Massey University},
            addressline={Private Bag 102-904 North Shore Mail Centre},
            city={Auckland},
            postcode={0745},
            country={New Zealand}}

\author[inst1]{Ian A. Bond}
\author[inst1]{Napoleon H. Reyes}




\begin{abstract}
Asteroids are an indelible part of most astronomical surveys though only a few surveys are dedicated to their detection. Over the years, high cadence microlensing surveys have amassed several terabytes of data while scanning primarily the Galactic Bulge and Magellanic Clouds for microlensing events and thus provide a treasure trove of opportunities for scientific data mining. In particular, numerous asteroids have been observed by visual inspection of selected images. This paper presents novel deep learning-based solutions for the recovery and discovery of asteroids in the microlensing data gathered by the MOA project. Asteroid tracklets can be clearly seen by combining all the observations on a given night and these tracklets inform the structure of the dataset. Known asteroids were identified within these composite images and used for creating the labelled datasets required for supervised learning. Several custom CNN models were developed to identify images with asteroid tracklets. Model ensembling was then employed to reduce the variance in the predictions as well as to improve the generalisation error, achieving a recall of 97.67\%.  Furthermore, the YOLOv4 object detector was trained to localize asteroid tracklets, achieving a mean Average Precision (mAP) of 90.97\%. These trained networks will be applied to 16 years of MOA archival data to find both known and unknown asteroids that have been observed by the survey over the years. The methodologies developed can be adapted for use by other surveys for asteroid recovery and discovery.
\end{abstract}



\end{frontmatter}


\section{Introduction}
\label{sec:intro}

Asteroids are among the millions of small bodies that inhabit our Solar System and are remnants from its formation. While popular sentiment most commonly associates asteroids with mass extinction events, the vast majority of asteroids pose no threat to us. The Main Asteroid Belt between Mars and Jupiter has the largest concentration of asteroids in our Solar System and this is where the majority of the asteroids seen in this research reside. Observing and tracking these small bodies gives us a better understanding of their complex orbital dynamics. Their composition and structure offer clues about the conditions when the terrestrial planets were formed 4.6 billion years ago.

Asteroids are part of the landscape of our night sky and appear in the imaging data of most astronomical surveys. However, as most surveys have a specialised purpose, their data is rarely mined for asteroids. Microlensing surveys like MOA  (\citet{Sumi2003}), OGLE (\citet{Udalski1993}), and KMTNet (\citet{Kim2016}) are particularly good for determining the rotation period and orbital trajectory of asteroids because they survey a given region of space several times each night (\citet{Gould2013}). This means that asteroids could spend several nights in the field of view of the telescope, giving us the opportunity to both observe their trajectory and analyse the light gathered from them. \citet{Cordwell2022} demonstrates the efficacy of extracting asteroids light curves from the MOA microlensing data.

Automated detection software has been part of surveys dedicated to discovering asteroids since the early 90s (\citet{Rabinowitz1991}). With improved computing power, other techniques for detecting moving astronomical sources such as shift and stack have also proven popular. In recent years, a leader in the field is the Pan-STARRS Moving Object Processing System or MOPS (\citet{Denneau2013}). Initially trained with simulated but realistic asteroid data for the Pan-STARRS telescopes, it takes transient candidates not associated with a known source and uses a complex tree-based spatial linking algorithm (\citet{Kubica2007}) to further parse and form associations between these point sources. MOPS does not work with imaging data but rather celestial coordinates, which reduces the computational cost. HeliolinC (\citet{Holman2018}) further improves on MOPS' efficiency with an approach that combines working with a heliocentric frame of reference and clustering sources that belong to the same object.

While these and other deterministic approaches have been successfully utilized for asteroid detection, applications of deep learning in the field remain in the early stages, potentially because of the lack of labelled data.  Deep learning offers the benefit of being able to learn representations directly from the raw data, making it a potentially valuable tool for asteroid discovery in archival astronomical data. The works that do apply deep leaning techniques note the benefits, particularly with greatly reducing the amount of data that must be examined by an astronomer, as we see next.
 
  \citet{Zoghbi2017} successfully applied both convolutional and recurrent architectures to reduce the amount of data to be vetted by astronomers looking for debris from long-period comets in the CAMS data\footnote{http://cams.seti.org/}. \citet{Lieu2018} applied neural networks to the task of detecting small solar system objects (SSO) in data simulated for the ESA's Euclid space telescope\footnote{https://sci.esa.int/web/euclid/}. They successfully used transfer learning and retrained three architectures from TensorFlow's Keras Applications library\footnote{https://keras.io/api/applications/} to distinguish between postage stamp cut-out images of asteroids and objects commonly mistaken for asteroids like cosmic rays, stars, and galaxies. \citet{ Duev2019} introduced DeepStreaks to aid in the ZTF's\footnote{https://www.ztf.caltech.edu/} quest for the discovery of near-Earth asteroids, which resemble streaks in the observations. Their model significantly reduced the number of candidate detections that had to be reviewed without sacrificing the detection sensitivity. \citet{Rabeendran2021} applied deep learning to the ATLAS\footnote{https://atlas.fallingstar.com/home.php} pipeline looking for near-Earth objects. It was successful in catching nearly 90\% of the false positive detections, thus greatly speeding up the process of followup observations. \citet{Duev2021} introduced Tails, which involved training an object detector to discover comets based on their distinctive morphology and it now forms a part of the ZTF's detection pipeline. Finally, \citet{Kruk2022} used deep learning to hunt for asteroid trails in archival data from the Hubble space telescope (HST)\footnote{https://hubblesite.org/}. They used composite HST images to make the asteroids trails longer and thus easier to detect. Their research also demonstrates the merits of citizen science for labelling the data and of mining archival data for asteroids with a deep learning-based toolkit.
 
We have seen that an important aspect of finding asteroids in survey data is isolating the sources that indicate a candidate object. In the case of microlensing data, this task is all the more challenging because of the extremely dense star fields observed, causing even the reference subtracted images to contain numerous spurious artefacts. Thus, rather than extracting the sources and then establishing connections between them, we propose a technique to instead enable extracting a cluster of sources that could represent part of the orbital arc of asteroids.

We present the methodology to create labelled datasets suitable for supervised learning, followed by convolutional neural network-based architectures for finding asteroids tracklets in microlensing surveys. Our classification models facilitate reducing the amount of data to be vetted and the object detection model localizes the potential tracklets within the classifiers' candidate detections. The bounding boxes predicted by the object detector could then be used to extract the potential sources, which in turn could be passed to orbiting link software, such as HeliolinC, to determine if they are valid solar-bound objects.

\section{MOA: Microlensing Observations in Astrophysics}
\label{sec:moa}
\begin{figure}
    \centering
    \includegraphics[width=0.48\textwidth]{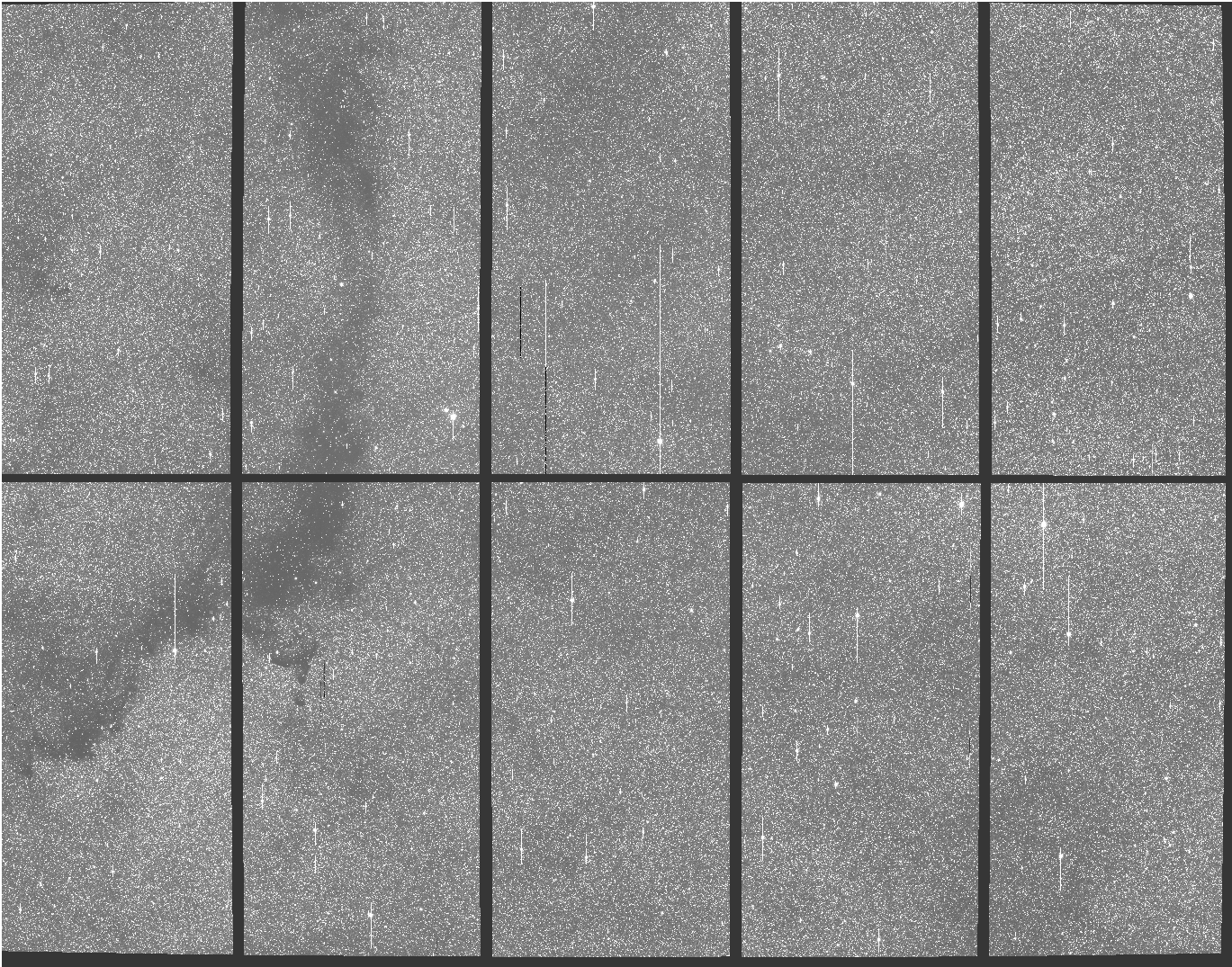}
    \caption{A single observation of the Galactic Bulge field, GB1, surveyed by MOA-II. North is right to left and east is bottom to top.}
    \label{fig:moa_gb1}
\end{figure}

This research utilizes archival data from the MOA (Microlensing Observations in Astrophysics) project, which is a Japan/New Zealand collaboration specialising in the search for gravitational microlensing events. They have been operating the 1.8m MOA-II optical research telescope at the Mt. John Observatory since 2004. It has a wide-field mosaic CCD camera called the MOA-cam3 (\citet{Sako2008}), which consists of 10 CCD chips. Each CCD chip is 3cm x 6cm and has 2048 x 4096 pixels, with a resolution of 0.01 arc minutes/ 0.6 arcseconds per pixel. Each exposure is 60 seconds long and all images are in a custom wide MOA-R band (630-1000 nm). The median FWHM seeing is ~1.7 arcseconds. The photometric precision is typically ~0.01 mag or better for I brighter than 16 and ~0.02 mag for I~18. The telescope has a total field of view of 1.7 x 1.4 square degrees. The survey field referred to as GB1 can be seen in its entirety in Figure \ref{fig:moa_gb1}. The CCD chips are numbered 1 to 10 clockwise, starting from the top-left.

MOA-II surveys the Galactic Bulge (GB) and the Large Magellanic Cloud (LMC), both of which are regions of the sky that are densely packed with stars. It operates at a high sampling rate, with some fields surveyed as often as every 10 minutes, which makes it particularly good for observing short duration events. The MOA image processing pipeline produces difference images (\citet{Tomaney1996, Alard1998, Alard2000, Bramich2008, Bond2001}, where a new observation image is subtracted from a reference image for the same region. The resultant difference image highlights the changes since the reference image was taken. These could either be transient astronomical phenomenon (like microlensing events or asteroids) or noise. The noise could be due to too-bright saturated stars, imperfect subtractions, satellite trails, instrumentation error, atmospheric dust, differential refraction, or proximity to a bright astronomical object (like the moon). As the fields surveyed by MOA-II are very dense, faint asteroids in particular are impossible to identify in the  original observation images. Thus, the MOA difference images form the basis of the datasets built for this research.  

\section{Building the Dataset}
\label{sec:dataset}
\begin{figure}[h!]
    \centering
    \includegraphics[width=.45\textwidth]{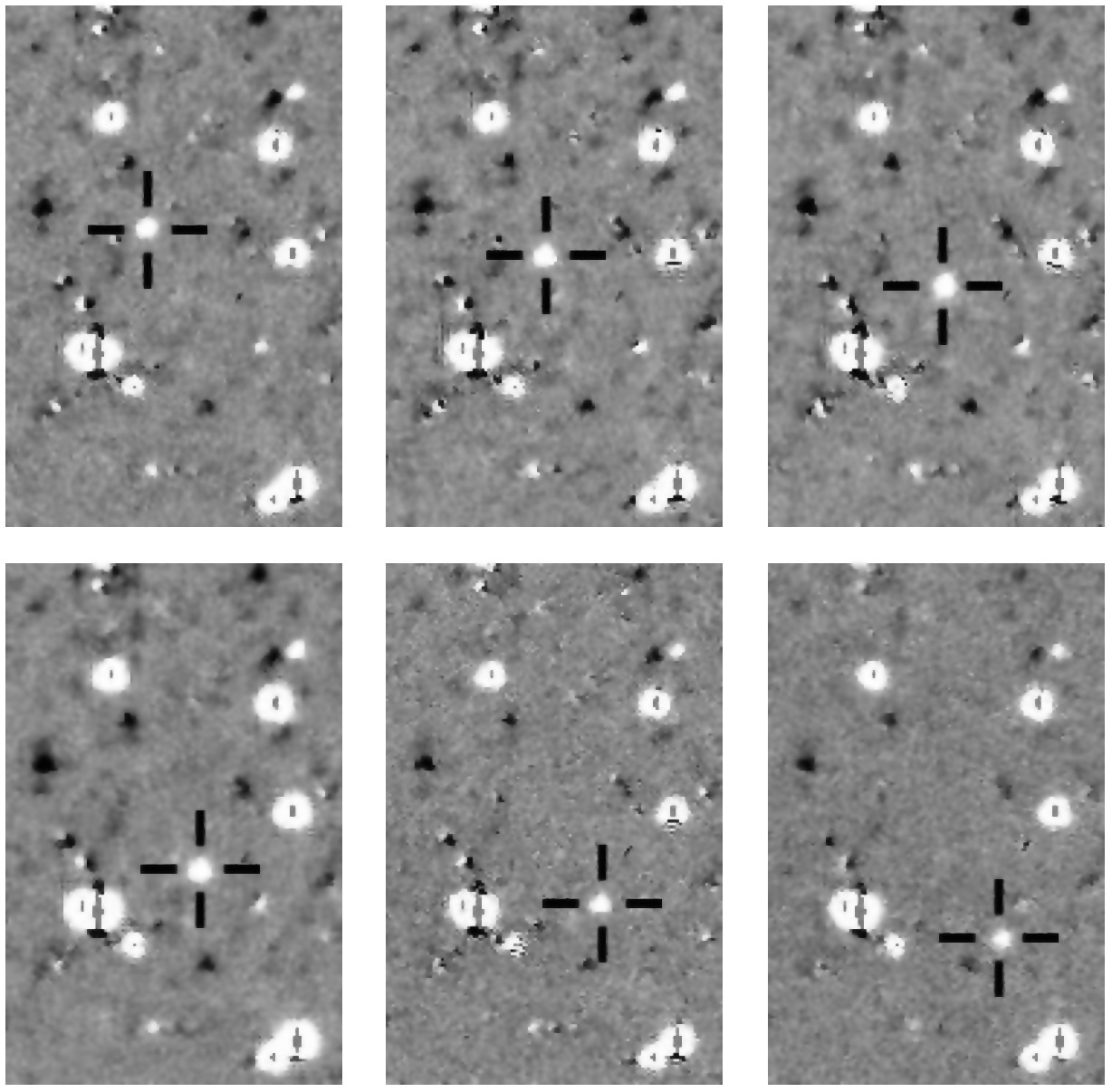}
    \caption{Six consecutive observations of the asteroid (78153) 2002 NX24 on 23-June-2006, spanning 70 x 108 arcseconds.}
    \label{fig:nx24obs}
\end{figure}

\begin{figure}[h]
    \centering
    \includegraphics[width=0.45\textwidth]{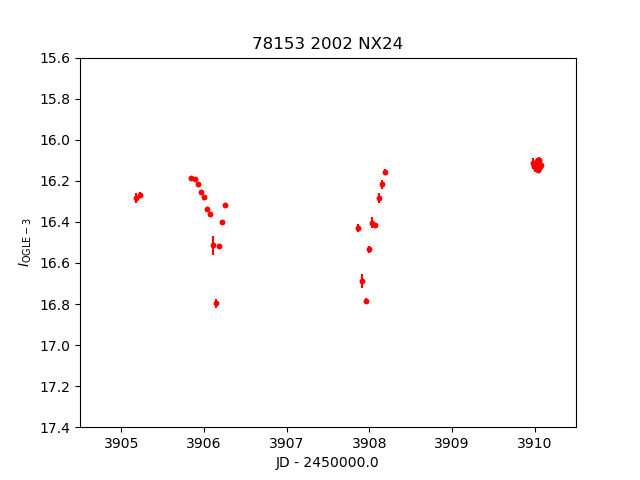}
    \caption{Light curve of asteroid (78153) 2002 NX24 extracted from the MOA-II data.}
    \label{fig:nx24lc}
\end{figure}

The Galactic Bulge is visible in the Southern Hemisphere between February and October. A combination of long nights and Bulge elevation in the mid-winter months make them the best time for observation. On clear nights with good seeing conditions, several of the MOA-II fields are surveyed frequently, at times as often as every 10 minutes. This provides ideal conditions for observing asteroids in our solar system. Figure \ref{fig:nx24obs} demonstrates this with six observations of a bright main-belt asteroid (78153) 2002 NX24, with a limiting magnitude (V) of 17.5, taken at 10-12 minute intervals on the 23rd of June, 2006. The light curve for this asteroid, extracted from the MOA-II data can be see in Figure \ref{fig:nx24lc}. The sky-plane velocities are about 0.5 arcseconds per minute, which are representative of the asteroids captured by MOA-II.

\begin{figure}[h]
    \centering
    \includegraphics[width=0.45\textwidth]{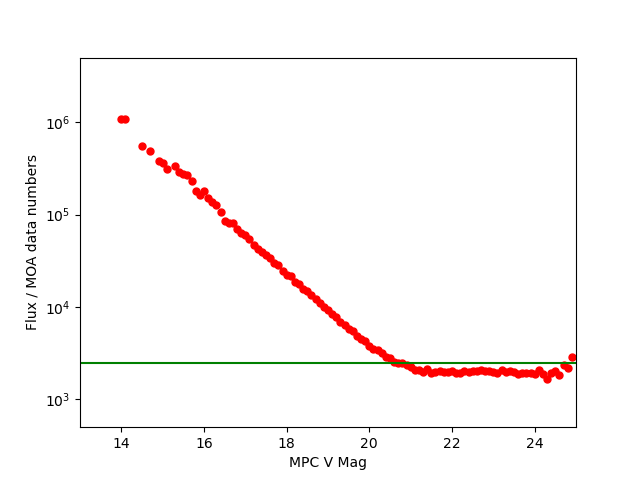}
    \caption{Measured MOA flux vs the MPC V magnitude for asteroids recovered from GB5-R5. The MPC value is reported to one decimal place and the results are binned in 0.1 magnitude bins. The red dots represents the median values of the fluxes. This starts to flatten out when V is around 21, at which point it is effectively the background being measured. The green line represents the working limiting flux commonly used in MOA microlensing work}
    \label{fig:fluxmag}
\end{figure}

The primary data used in this research consists of all observations from 2006 - 2019 from one chip (chip 5) in the CCD array for the field GB5, amounting to 49,901 difference images (GB5-R5). This field is surveyed very frequently and the cadence of most observations is between 5 and 20 minutes. The Minor Planet Center (MPC) was queried\footnote{https://www.minorplanetcenter.net/cgi-bin/checkmp.cgi} via a screen scraper to get all known asteroids expected to traverse through the region encompassed by GB5-R5. To determine the limiting magnitude of our survey, we have compared the fluxes measured at the positions of recovered asteroids from GB5-R5 with the V magnitudes provided by MPC. The results are shown Figure \ref{fig:fluxmag}, where the data has been binned by MPC V magnitude with a bin size of 0.1 mag with the median flux values calculated for each bin. For microlensing measurements, MOA adopts a working value of 2500 for the limiting data number. This corresponds to a MPC V magnitude ~20.5 which we adopt as our limiting magnitude here.

\begin{figure}[hb]
    \centering
    \includegraphics[width=0.45\textwidth]{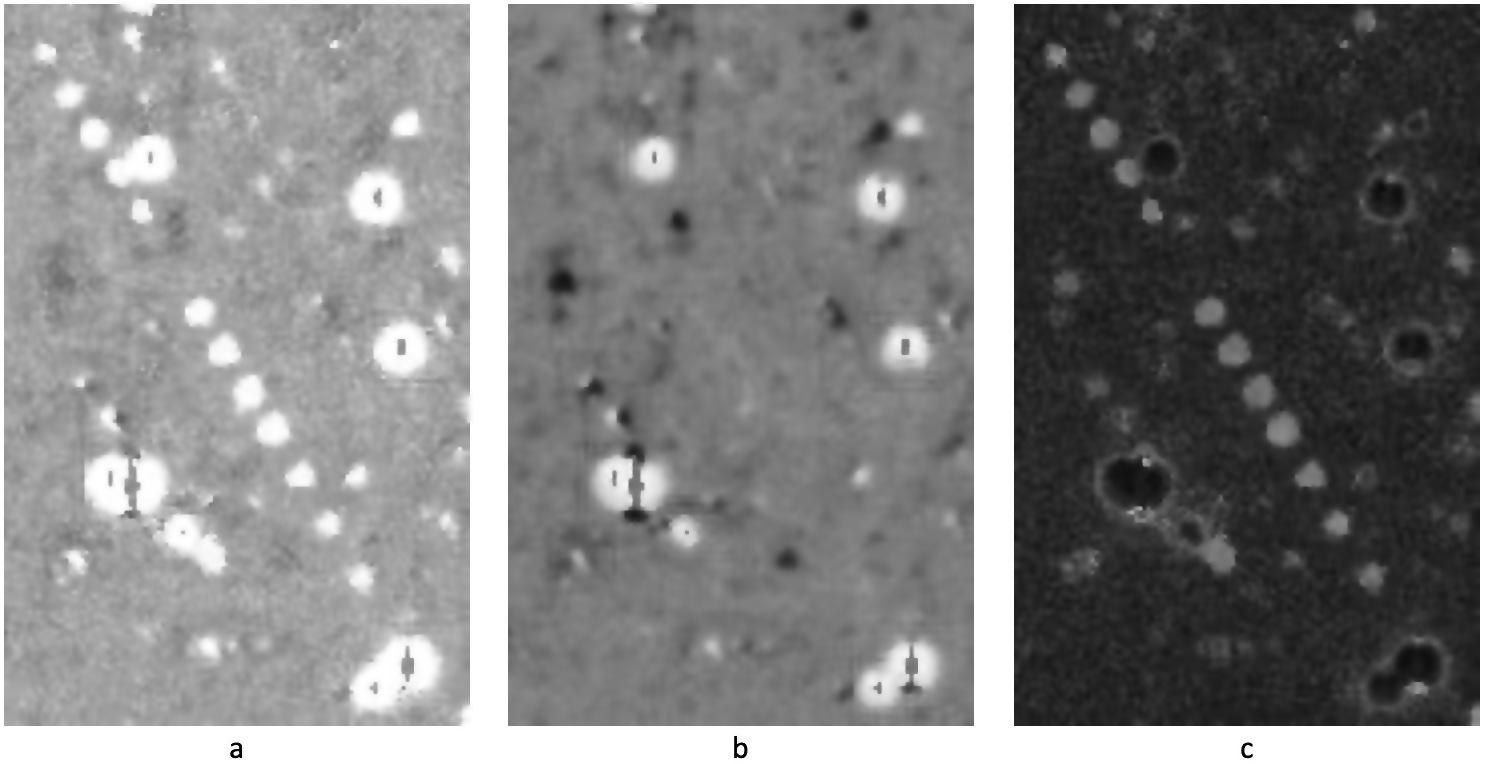}
    \caption{Stacking all of a night's observations (23-June-2006) on one chip in one field gives us a clear tracklet for asteroid (78153) 2002 NX24. In (a) the observations are stacked by brightest pixel; in (b) they are stacked by the median pixel; and in (c) we see the result of subtracting (b) from (a).}
    \label{fig:nx24stacks}
\end{figure}

Since asteroids move appreciably with respect to the background stars in each consecutive exposure, asteroid tracklets can be clearly seen by stacking nightly observations. We define a tracklet as part of the orbital arc of the asteroid as it travels through the field of view of the telescope. We started with a stack composed of the brightest pixels. While the tracklets were visible, so was the bright background and the over-bright stars (Figure \ref{fig:nx24stacks}(a)). To highlight the tracklets better, the stack image was further simplified by subtracting the brightest pixel stack from the median pixel stack. This gave us an image without the bright background and saturated stars (Figure \ref{fig:nx24stacks}(c)), leaving behind the noise and moving objects. Only nights with 3 or more observations were considered when generating the subtracted stacks. The GB5-R5 dataset resulted in 2252 subtracted stack images within which the search for asteroid tracklets commenced.

\begin{figure}[!ht]
    \centering
    \includegraphics[width=0.48\textwidth]{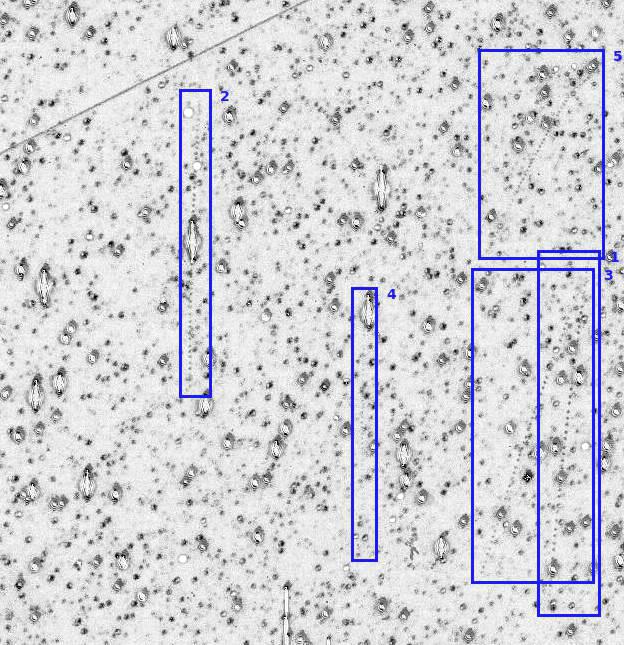}
    \caption{A stack of all 51 observations from the night of 15-May-2008 reveals 5 asteroid tracklets clearly visible in a sub-region of the stack image. Numbered in order of appearance, these are: 1 - (103842) 2000 DQ33 (19.5); 2 - (148657) 2001 SX124 (19.6); 3 - (152083) 2004 RH30 (19.9); 4 - (582743) 2016 AT221 (20.4); 5 - (338789) 2005 SZ154 (20.4). The line/streak on the top left is from a satellite. This image has been inverted and brightened to improve visibility (original in Figure \ref{fig:quintet_apdx}).}
    \label{fig:quintet}
\end{figure}

The ephemeris data from the MPC, together with the astrometric calibrations specific to MOA-II, were used to extrapolate the minimum and maximum (x, y) positions for asteroid tracklets in subtracted stack images. This was used to crop sub-regions expected to contain tracklets from each stack. Asteroids that fell beyond the boundaries of a CCD chip were ignored. Further, by visual inspection of the tracklets, it was established that only asteroids with a limiting magnitude of 20.5 or brighter are visible in the MOA-II exposures. This is also consistent with what we see in Figure \ref{fig:fluxmag}, where the median flux around V~21 flattens out and is akin to measuring the background. The objects with magnitude between 19.5 and 20.5 are often at the very edge of visibility, requiring excellent seeing condition and a high signal to noise ratio to be visible in the observations. Careful examination of the images resulted in 2073 tracklets from 1178 distinct asteroids over 1078 distinct nights. Figure \ref{fig:quintet} (374 x 387 arcseconds) displays some of the tracklets captured in the observations on the night of 15-May-2008. Some of the tracklets are very faint.

Tracklets come in a variety of sizes, but we need images of a uniform size for training neural networks. The original size of the observations was deemed too big to be used as is because of the computational cost as well as the potential for the tracklets to be lost amongst the other artefacts in the image. Therefore, a decision was made to split each image into 128 x 128 tiles, giving us 512 tiles per image. Each 128x128 tile spans 76 x 76 arcseconds. For GB5-R5, this gave us a total of 551,936 images from the 1078 nightly stacks that contained visible asteroid tracklets. 

\begin{figure}[!ht]
    \centering
    \includegraphics[width=0.48\textwidth]{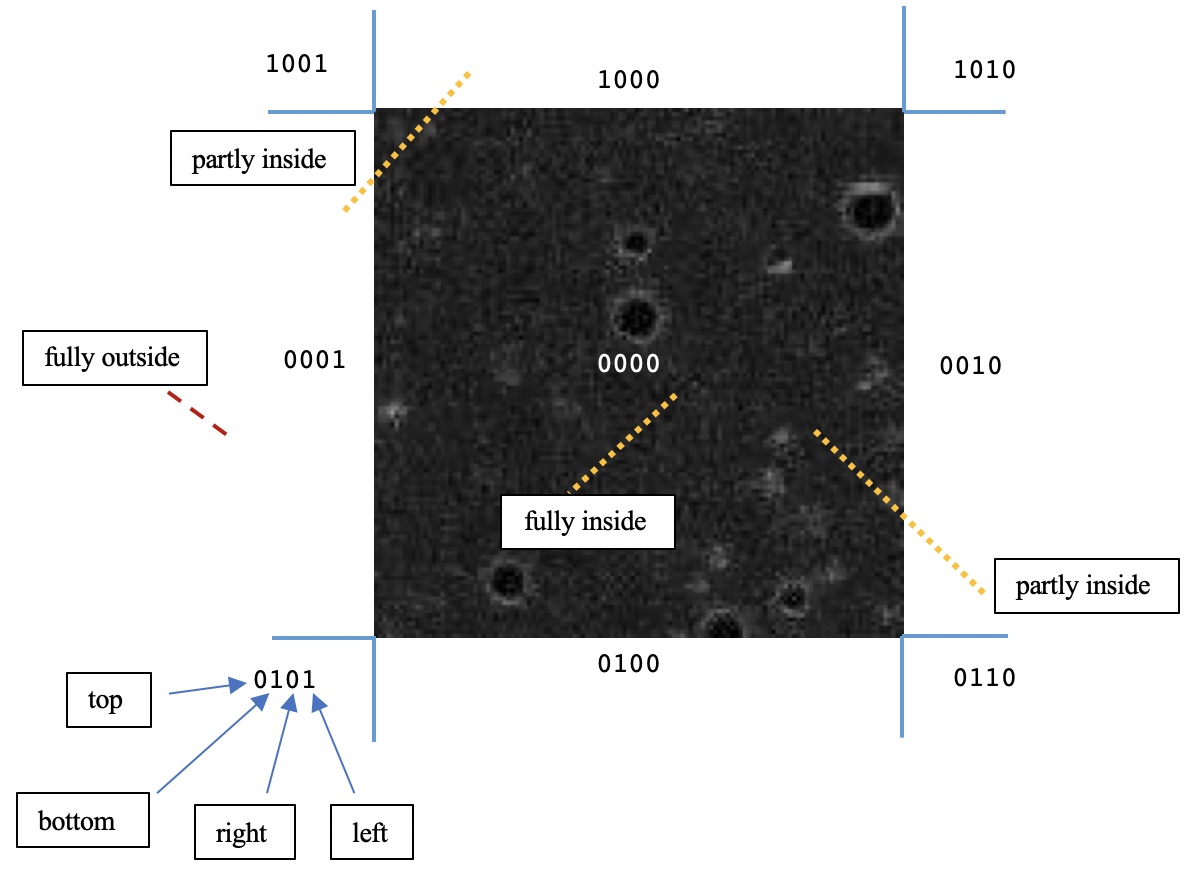}
    \caption{The Cohen-Sutherland line clipping algorithm was used to reject tracklets outside the area of interest and determine intersection points of the ones partially inside.}
    \label{fig:cohen}
\end{figure}

The Cohen-Sutherland line clipping algorithm (\citet{cohensutherland1973}) was next used to locate the asteroid tracklets within these images. The algorithm involves dividing a rectangular space (in this case, an image) into nine regions - eight "outside" and one "inside", as seen in Figure \ref{fig:cohen} - and determining which of the lines (tracklets) are fully or partially inside  the area of interest (the image). Each of the nine regions have associated outcodes (4-bit numbers) that are calculated by performing bitwise operations after comparing the start and end points of the line/tracklet with the coordinates of the image. For example, the outcode 0001 indicates that the end point is center left and the outcode 1000 indicates that the end point is center top. A bitwise OR of these two codes returns 1001, which is a non-zero value, and the bitwise AND returns 0000, which is zero, which in turn indicates that the tracklet is partially inside the image. Thus, there are three possible solutions for any line:
\begin{itemize}
    \item If both endpoints of the line are inside the area of interest, the bitwise OR computation returns 0 (trivial accept).
    \item If both endpoints of the line are outside the area of interest, they will share at least one outside region and the bitwise AND computation returns a non 0 value (trivial reject). 
    \item If both endpoints are in different regions, at least one endpoint will be outside the image tile. In this case, the intersection point of the tracklet's outside point and image tile boundary becomes the new endpoint for the tracklet and the algorithm repeats until the bitwise operation returns a trivial accept or reject.
\end{itemize}

\begin{figure*}
    \centering
    \includegraphics[width=\textwidth]{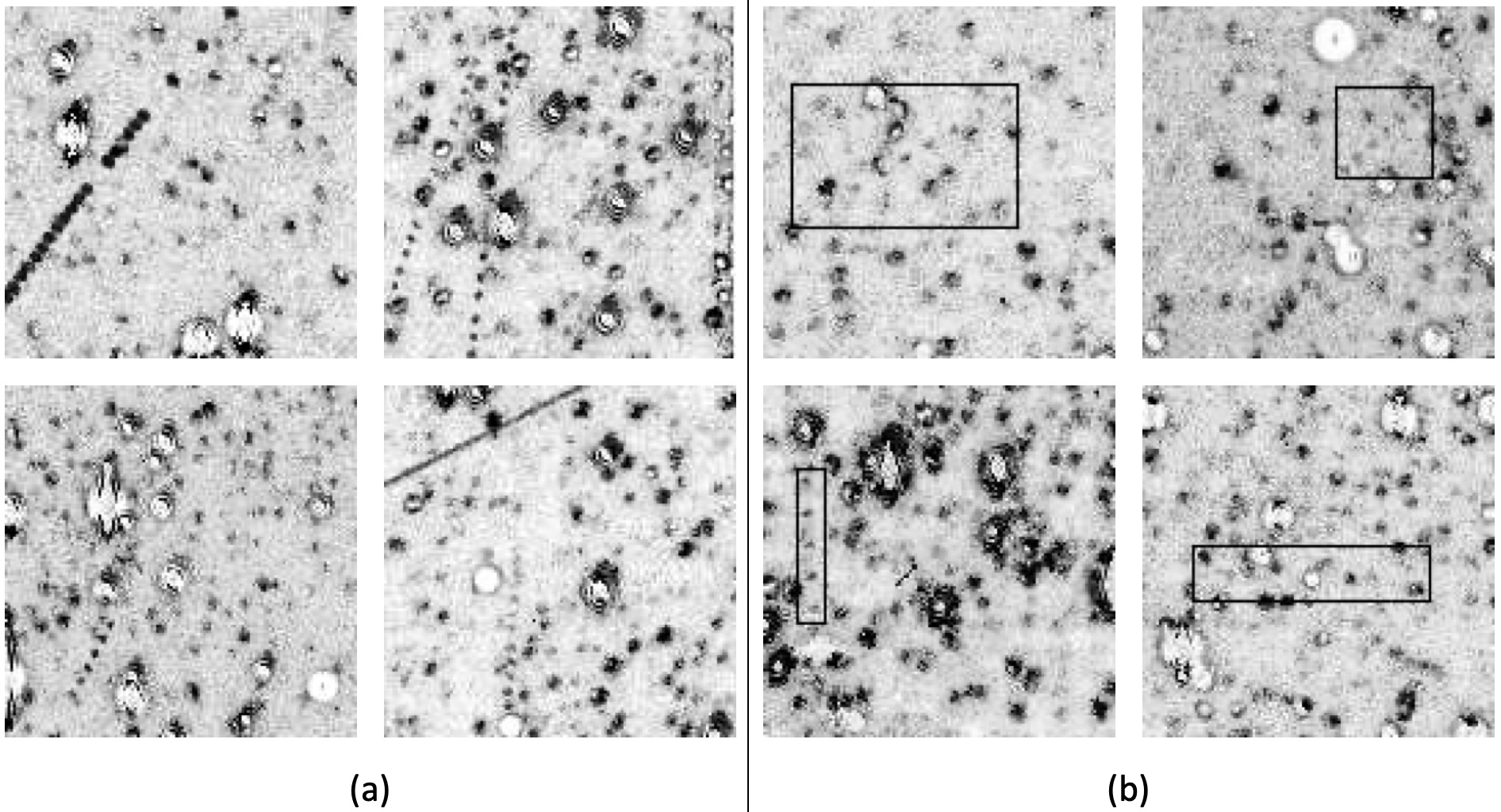}
    \caption{Asteroid tracklets seen in 128x128 sub-regions from subtracted stack images. In (b), the tracklets are localized with bounding boxes. Asteroids in the top row, left to right: (38102) 1999 JM18 (18.4); (152083) 2004 RH30 (19.9); (103842) 2000 DQ33 (19.5); (283261) 2011 FR142 (20.0); and (375674) 2009 HD5 (19.8). Asteroids in the bottom row, from left to right: (48617) 1995 HR2 (19.5); (148657) 2001 SX124 (19.6); (97948) 2000 QF124 (19.6); and (74978) 1999 TY234 (18.7). Note that these images have been inverted and brightened to improve visibility (original in Figure \ref{fig:tracklets_apdx}). }
    \label{fig:tracklets}
\end{figure*}

The application of this algorithm gave us 8341 tiles that potentially had visible tracklets and 543,595 without known asteroids. Once again these were meticulously scanned to ensure that a tracklet could clearly be seen. Any tile with less than 3 points of a tracklet were rejected, along with tiles that did not contain a visible portion of a tracklet. At the conclusion of this process, there were 4153 128 x 128 tiles with visible tracklets from GB5-R5. Some of these can be seen in Figure \ref{fig:tracklets}(a).

For classification, the images where the tracklet was too obscured by noise or where there were less than three clearly visible adjacent point sources related to a tracklet were removed. Following this, there were 4072 images with tracklets, which were split into the training/ validation/ test set with 3322/ 415/ 335 in each, respectively. The images in the training and validation set were further augmented by rotating 180 degrees, flipping horizontally and vertically, brightening, darkening, blurring, and both increasing and decreasing the contrast. There was no overlap between the training, validation, and test sets. A further test set composed of all the observations on a single night from all ten chips in fields GB3, GB4, GB5, GB9, GB10, and GB14 was also constructed (GB-All), and it had 300 tracklet and 2000 no-tracklet images. The purpose of the GB-All test set was to evaluate how the networks performed with data they had never seen; as before, the tracklets were from observations with a maximum cadence of 20 minutes and there are a minimum of 3 sources per image. The classification networks were also trained to recognize images with no tracklets. Forty images without known tracklets were randomly chosen from each of the 512 sub-regions, giving us 20,480 images. This collection was also visually inspected and any images that could potentially have tracklets were removed. This process resulted in 19,682 images, which were split into the training/validation/test sets with 15,595/2039/2048 images, respectively. The same set of augments was generated for the no-tracklet training and validation data, but since this set was already significantly larger, only a random 35\% of the no-tracklet augments were used. Note that this distribution does not reflect the true ratio of tracklet vs no-tracklet images; realistically, on a good night, we might expect as many 10 tracklet images for every 100 no-tracklet images. The roughly 1:5 ratio here serves to better focus the network to learn the tracklet pattern while providing a suitable number of contrasting no-tracklet images.

As the asteroid tracklets are a well-defined pattern of blobs in an image, we further considered training a deep-learning based object detector to localize them in the images. This would facilitate locating the tracklets that were faint and/ or obscured by noise as well as distinguish the area of interest in the images. We utilize YOLOv4 (\citet{Bochkovskiy2020}), which requires the objects of interest - in our case, tracklets - to be enclosed in bounding boxes, with the coordinates of the object's centroid saved with respect to the dimensions of the image. As a result of applying the Cohen-Sutherland line clipping algorithm, we had the probable intersection/end points of tracklets within the 128 x 128 images. These intersection points were used to extrapolate bounding boxes for tracklets, thus highlighting them in the images. Further visual inspection and manual adjustment was undertaken to ensure that bounding boxes encapsulated the tracklets correctly without any superfluous background included (Figure \ref{fig:tracklets}(b)). All 4153 tracklet images were used, with a training/ test split of 3737 and 416, respectively. 

\section{Deep Learning Framework}
\label{sec:cnn}
The appeal of neural networks and deep learning is in their capacity to make predictions about complex data in real time after they have been trained with a representative dataset. In the last decade, deep learning has matured significantly and has proven to be highly effective for classification and object detection tasks. Here, we discuss training several convolutional neural network (CNN) based architectures to find asteroids tracklets in the subtracted stack images. 

The proposed model consists of an ensemble of five classifiers that produces a probability that a 128x128 composite image tile contains a tracklet. Tiles believed to contain tracklets are then passed to YOLOv4, which has been trained to localize tracklets by adding a bounding box around the tracklet within the image.

\subsection{Classification}
\label{sec:classification}
A convolutional neural network or CNN is a network architecture that is designed to take advantage of the 2D structure of an input image. Using a series of convolutional layers that utilise filters, implemented using local connections and shared weights, it can extract meaningful features directly from data, thereby eliminating the need for manual feature extraction.

Traditionally, the number of filters in a layer increases with the depth of the network. The received wisdom is that the more layers a network has, the better it is at extracting more complex features from an image, and thus learning from complex data. The limiting factors here, however, are the size of the input and how localised the features we are interested in are, as well as the amount of available training data. After a given number of layers, a network could start overfitting to the data by focusing on the irregularities in the images. So, while increasing the number of convolutional layers improves the performance of the network, it is not the case that the deeper network is always the best option. Part of the challenge is finding the ideal depth for a network to optimally predict the output.  Simonyan et al. introduced VGGNet (\cite{Simonyan2014}) and investigated the effects of increasing the depth of convolutional layers on the model's classification and localisation accuracy in the large-scale image recognition setting (ILSVRC challenge).  VGGNet has a compelling simplicity in its architecture; it uses a stack of consecutive convolutional layers to reduce the number of parameters, leading to faster convergence and reducing the overfitting problem. GoogLeNet (\cite{Szegedy2015}), popularly known as Inception, introduced the idea of modifying the width of a network as well as employing filters of a variety of sizes to better capture multi-scale data from images.  Relative to the VGGNet architecture, it is able to reduce the number of parameters and computational cost further. Residual networks (\cite{He2016}), or ResNets, were introduced to address the vanishing gradient problem as networks go deeper by presenting an alternative pathway for the algorithm to follow, called the skip connection. The central element in this architecture is the residual block, which consists of two convolutional layers with 3x3 filters. The input of this is added to the output of the second convolution, thus creating a shortcut connection.

In our work, a variety of CNN classification architectures were trialled to determine the best combination of filters and layers to suit our purpose. After considerable experimentation with well-established architectures such as VGG-16 and VGG-19 (\cite{Simonyan2014}), Inception (\cite{Szegedy2015}), and ResNet50 (\cite{He2016}), various custom models were constructed to determine if better performance could be obtained. Table \ref{tbl:cnn} describes the structure of five of these custom architectures, three of which are VGG-like (MOA-12, MOA-14, MOA-15), and two are composed of hybrid Inception-ResNet modules (Hybrid A, Hybrid B). All of the custom architectures have significantly fewer parameters than their established counterparts.  It is our finding that having an over-parameterised deep network architecture causes over-fitting of the training data, and so we reduced the number of layers and filters in the network to reduce the number of neurons and weights involved.

\begin{figure}[!ht]
    \centering
    \includegraphics[width=0.45\textwidth]{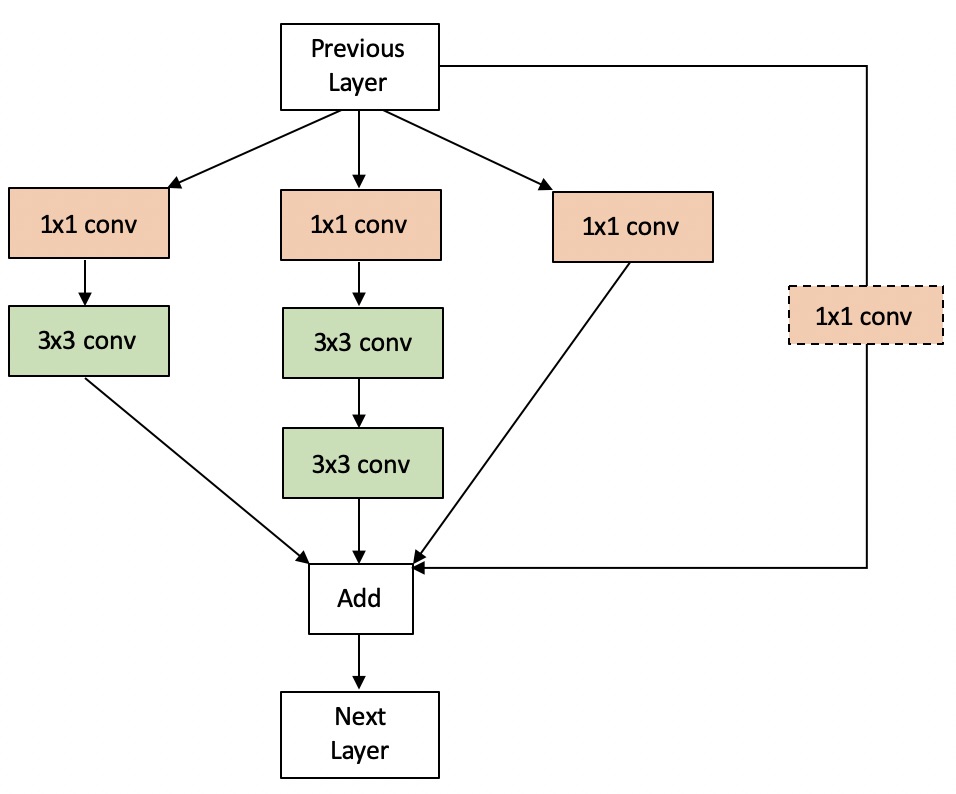}
    \caption{Hybrid module combining salient features of a ResNet block and an Inception module}
    \label{fig:hybrid}
\end{figure}

\begin{table}[!ht]
    \centering
    \includegraphics[width=.5\textwidth]{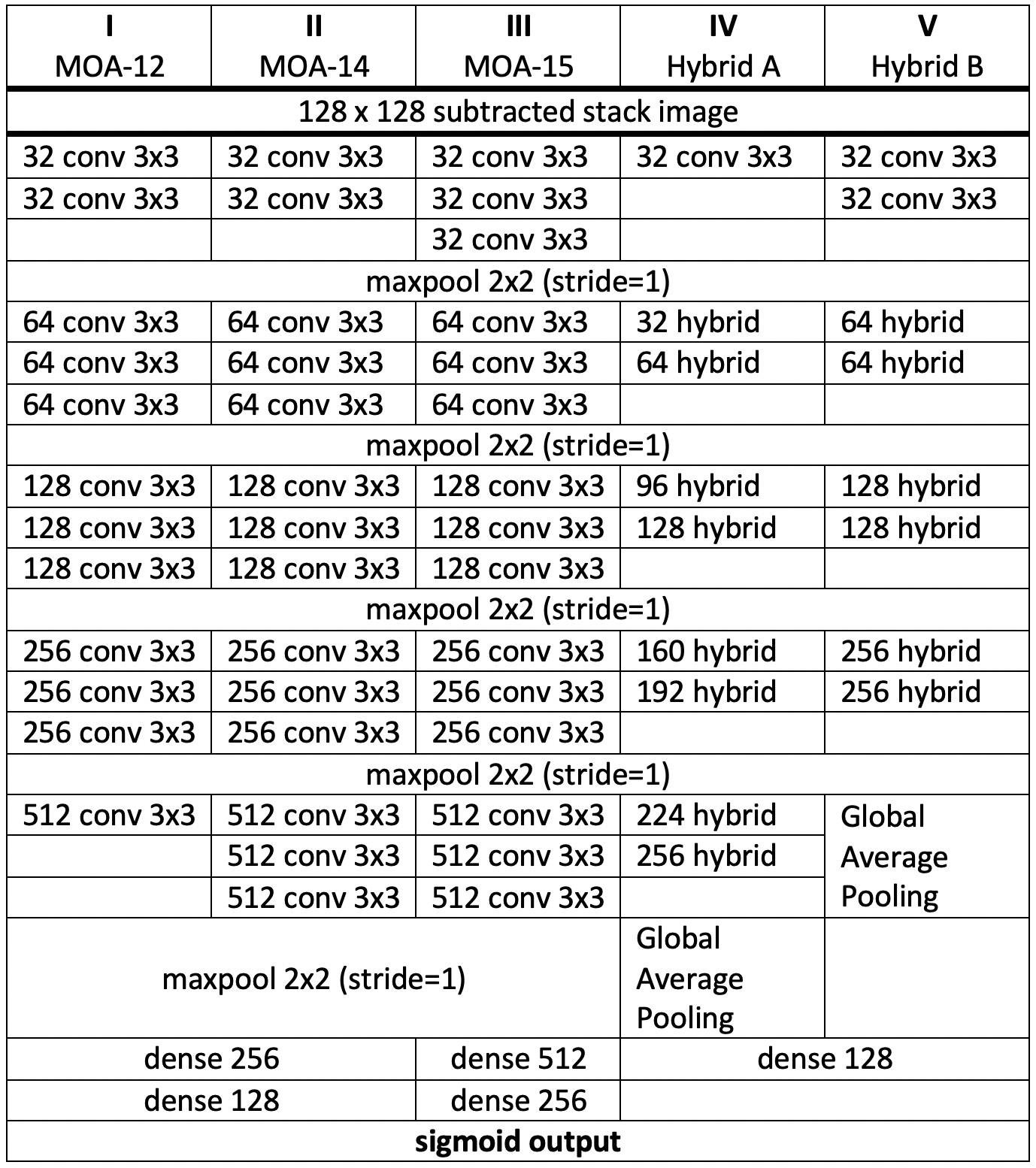}
    \caption{Configuration of custom classification architectures MOA-12, MOA-14, MOA-15, Hybrid A, and Hybrid B. ReLU activation is used in each layer and dropout is used after each fully connected layer.}
    \label{tbl:cnn}
\end{table}

The hybrid module (Figure \ref{fig:hybrid}) consists of four branches composed of convolutional layers, each of which use the ReLU activation (\citet{Glorot2011}; \citet{Nair2010}) and are combined with the 'add' function. The final branch carries the input to the 'add' function, with the convolution being used only when the number of filters in the previous layer are not equal to the current layer's filters. The many 1x1 convolutions in the hybrid module may seem superfluous but removing even one negatively affects the performance of the network. This may be because of the beneficial complexity introduced by the associated non-linearity. 

All of the classification architectures were trained on a Linux machine running Ubuntu 18.04 with a NVIDIA Quadro M4000 GPU (8 GB, 2.5 TFLOPS). The code was written in Python 3.6 in the Jupyter Notebook environment. TensorFlow GPU 2.4.1 (CUDA 11.0), along with the Keras deep learning API, was used for creating the custom CNN models tested. The Keras Applications implementations were used for the established models. A batch size of 32 was used for training each network model. Training was set to run for 50 epochs, with a callback for early stopping if the validation loss failed to minimise after a set number of epochs. The Adam (\citet{Kingma2014}) optimiser was used and the loss function minimised was binary cross entropy. The learning rate was initialised at 0.0001 for all networks except Hybrid A, which started with a learning rate of 0.001. In each case, the learning rate was reduced after 15 epochs and at scheduled intervals after that point. Callbacks were included to save the best weights for both validation and training accuracy. The established networks performed better when pre-loaded with ImageNet weights before fine-tuning with the MOA-II data. MOA-12, 14, and 15 took on average 3 hours to train and the Hybrid models, 6 hours.

\subsection{Object Detection}
\label{sec:yolo}

\begin{figure*}
    \centering
    \includegraphics[width=\textwidth]{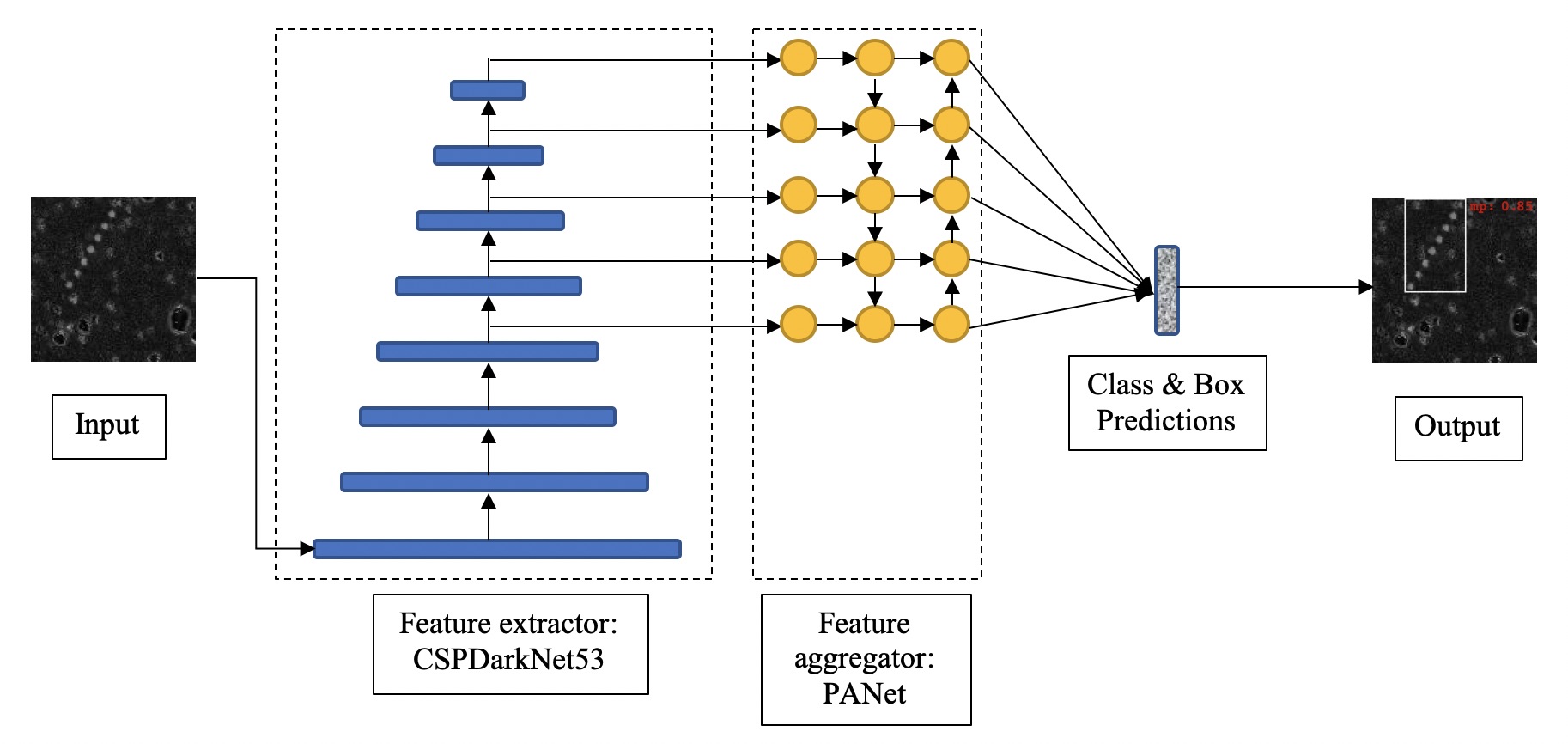}
    \caption{Architecture of YOLOv4: a feature extractor that is composed of a 53-layer DenseNet with cross-stage partial connections and spacial pyramid pooling along with a feature aggregator that effectively combines the feature maps from the higher and lower resolution layers.}
    \label{fig:yolo_model}
\end{figure*}
In recent years, CNNs have been taken beyond classifying images and have been applied to object detection and localization in images. Here, we utilize the YOLOv4 object detection architecture for localizing asteroid tracklets in our tile images.

The aim of the original YOLO (You Only Look Once) object detection architecture (\citet{Redmon2016}) was to make object detection both fast and accessible. As a single-shot architecture, it was capable of making class and bounding box predictions with the feature maps produced by a single CNN network. The simplicity of the feature extraction CNN at the heart of YOLO was in direct contrast to its complex loss function, which computed the classification loss, localization loss, as well as the loss quantifying the network's confidence in the prediction. Later versions of the network introduced default anchor boxes, which are predefined bounding boxes that are adjusted and refined during training to encompass the objects of interest in an image(\citet{Redmon2017}), and more complex feature extractors to make predictions on multiple scales (\citet{Redmon2018}). 

This research applied YOLOv4 (\citet{Bochkovskiy2020}), which is the latest evolution of the architecture with a major overhaul to include several new techniques, making the model state-of-the-art while still being easy to train. In particular, YOLOv4 ensures that the lower level features are propagated through both the feature extractor as well as the feature aggregator. Figure \ref{fig:yolo_model} illustrates the architecture of YOLOv4. The convolutional backbone feature extractor for the architecture is composed of a 53 layer DenseNet (\citet{Huang2016}) with the cross-stage-partial (CSP) connections of CSPNet (\citet{Wang2020a}). DenseNets extend ResNet's concept of skip connections by adding connections between all the layers in the network in a feed-forwards fashion. Feature maps from all preceding layers are concatenated and form the input for any given layer, ensuring that low-level features are propagated through the network. CSP connections involve splitting the input feature map into two parts, one of which goes through the dense block and the other goes straight through to the next transitional step. Additionally, the network includes spatial pyramid pooling (SPP) (\citet{He2014}) after the last convolutional layer. This has the effect of separating out the most important features and increasing the receptive field. The final feature map is divided into \(m\ \times\ m\) bins, following which maxpooling is applied to each bin. The resulting feature maps are concatenated and represent the output of the feature extractor.  

CNNs naturally attain a pyramid-like structure with each layer as the image goes from high to low resolution.  As we get deeper in a CNN, we lose the fine-grained details of the input, which usually makes it harder to detect small objects. As the resolution lowers, however, the filters learn ever more complex abstractions about the image, making the feature maps more semantically rich. Therefore, is it desirable to combine the feature maps from the higher resolution layer with the more semantically rich ones to facilitate detecting objects at multiple scales. This task falls to a feature aggregator and YOLOv4 uses the approach suggested by PANet (\citet{Liu2018}). The feature maps are concatenated from both the top-down and the bottom-up path, ensuring the propagation of semantically rich localization information through to the final part of the network where the class probability and bounding box predictions are made. Each predicted bounding box consists of five elements: centre-x, centre-y, width, height, and confidence. The (\(centre-x, centre-y)\) coordinates are relative to the dimensions of the predicted box; the width and height are relative to the whole image. The confidence score represents the likelihood that the cell contains the object as well as how confident the model is about its predictions.

YOLOv4 also updates the loss function to include Complete Intersection over Union (CIoU) loss (\citet{Zheng2020}) to train the network to effectively determine the direction in which to shift the weights to better match the labelled bounding boxes. 
Finally, the model also includes a raft of new additions: a new activation function, MISH (\citet{Misra2019}) that provides smoother gradients; updates to the spatial attention module (\citet{Woo2018}) and multi-input weighted residual connections (\citet{Tan2020a}) to better suit the architecture; and new data augmentation techniques, Mosaic and self-adversarial training. 

The YOLO family of models are supported by Darknet (\citet{Redmon}), a custom framework written in C and CUDA and designed for fast object detection. The Darknet implementation of YOLOv4 was trained with the MOA-II dataset via the Google Colaboratory with a hosted GPU runtime environment. The model was first trained with the default anchor boxes, before k-means clustering was used to discover anchor boxes that might prove better suited to finding tracklets in astronomical data. After several variations were tested, the best combination of anchor boxes was discovered by hand-engineering the various clusters. The learning rate of 0.001 was used along with a batch size of 32, with 8 subdivisions, 6000 max\_batches, and with steps set to 4800,5400.

\section{Results}
\label{sec:results}
To evaluate the classification networks, we rely on the metrics derived from a confusion matrix. A confusion matrix breaks down the predictions made by a classifier into 4 outcomes: True Positive (TP), True Negative (TN), False Positive (FP), and False Negative (FN). The "positive" cases are where images were classified as containing tracklets and the "negative" cases are where no tracklets were detected. In this case, the ideal scenarios would be to have as few false negatives as possible, together with a manageable number of false positives. Here, the networks are evaluated based on their recall, F2 score (the weighted harmonic mean of the precision and recall), and the PR AUC (area under the precision-recall curve). 

\begin{table}[!ht]
    \centering
    \includegraphics[width=0.48\textwidth]{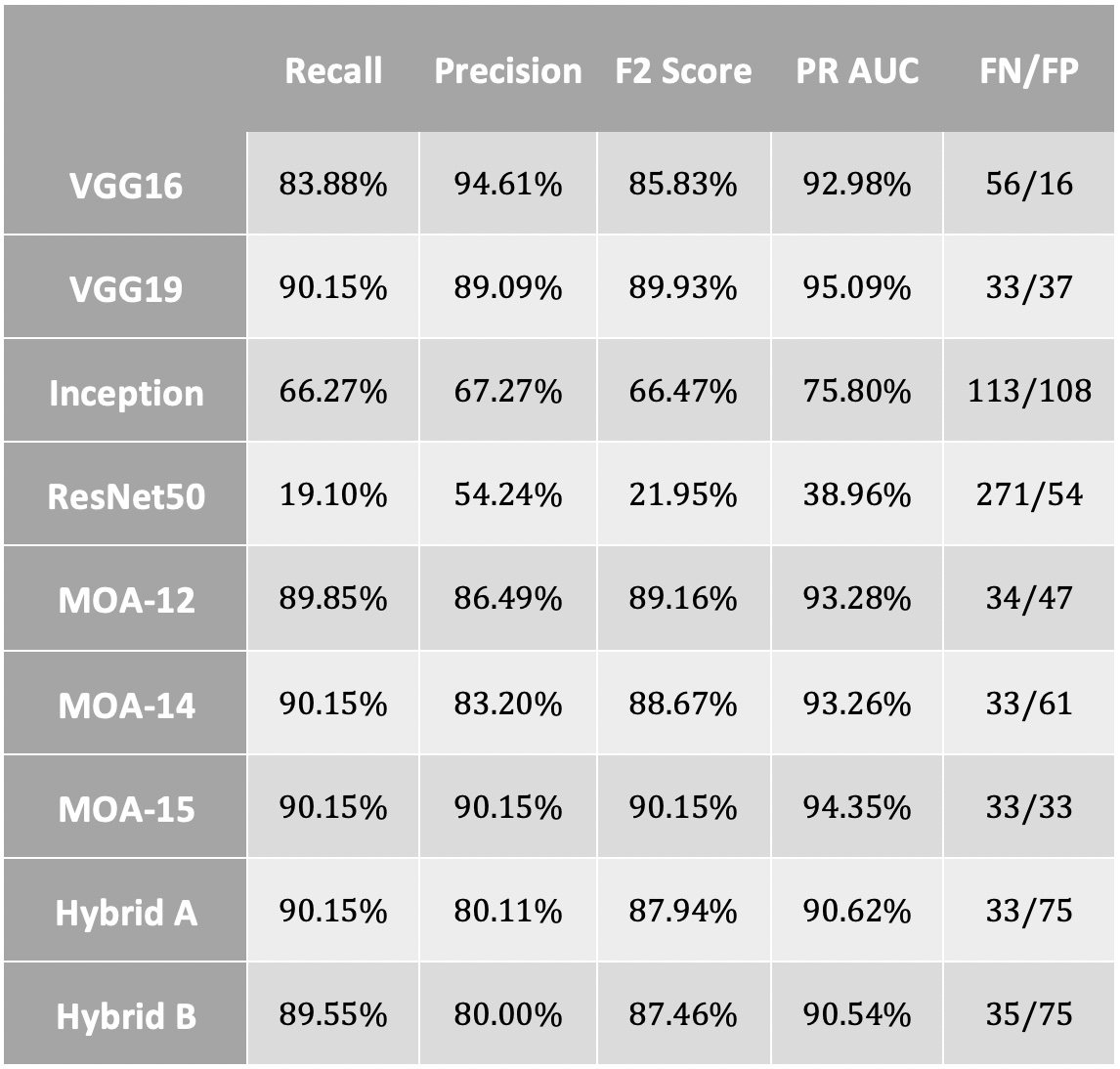}
    \caption{Evaluation metrics for the GB5-R5 test set taken at probability threshold 0.5}
    \label{tbl:cnn_gb5}
\end{table}

The evaluation metrics for each classifier for the GB5-R5 test set are in Table \ref{tbl:cnn_gb5} and for the GB-All (28-06-2013) test set are in Table \ref{tbl:cnn_gball}. All of the metrics were taken at the 0.5 confidence threshold, with values over 0.5 indicating the presence of an asteroid tracklet in the image.  Reviewing a combination of the PR AUC, F2 Score, and Recall, we can see the custom networks generalised well when making prediction about data from fields and chips they had never seen. Rather than selecting a single network from among these, all five custom classifiers were configured as an ensemble. Each network makes predictions about an input image and two approaches were trialled for selecting the winning prediction: averaging the predictions from all five classifiers (avg) or selecting the highest predicted value (max). 

\begin{table}[!ht]
    \centering
    \includegraphics[width=0.48\textwidth]{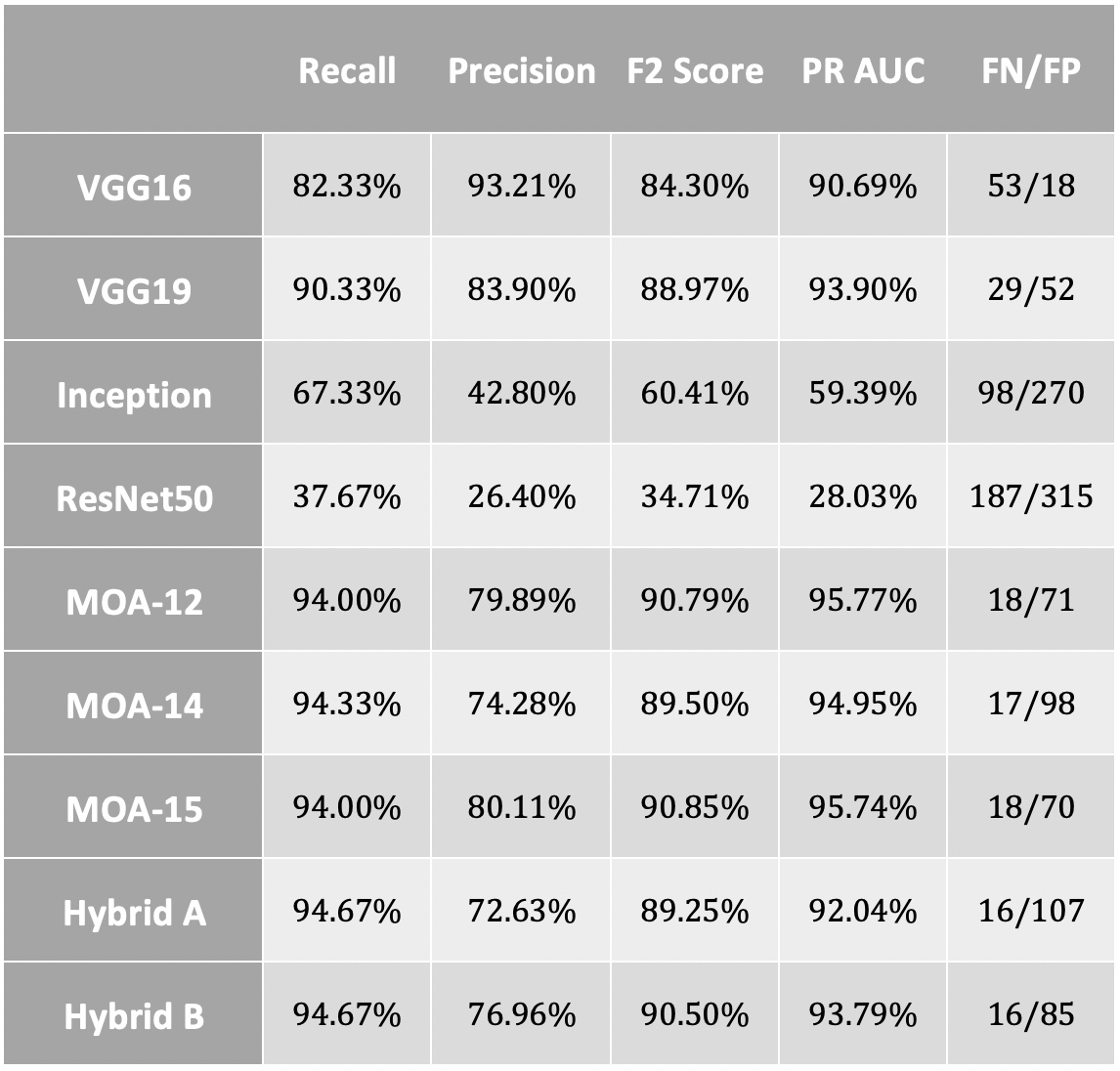}
    \caption{Evaluation metrics for the GB-ALL test set taken at probability threshold 0.5}
    \label{tbl:cnn_gball}
\end{table}

\begin{figure}[!ht]
    \centering
    \includegraphics[width=0.48\textwidth]{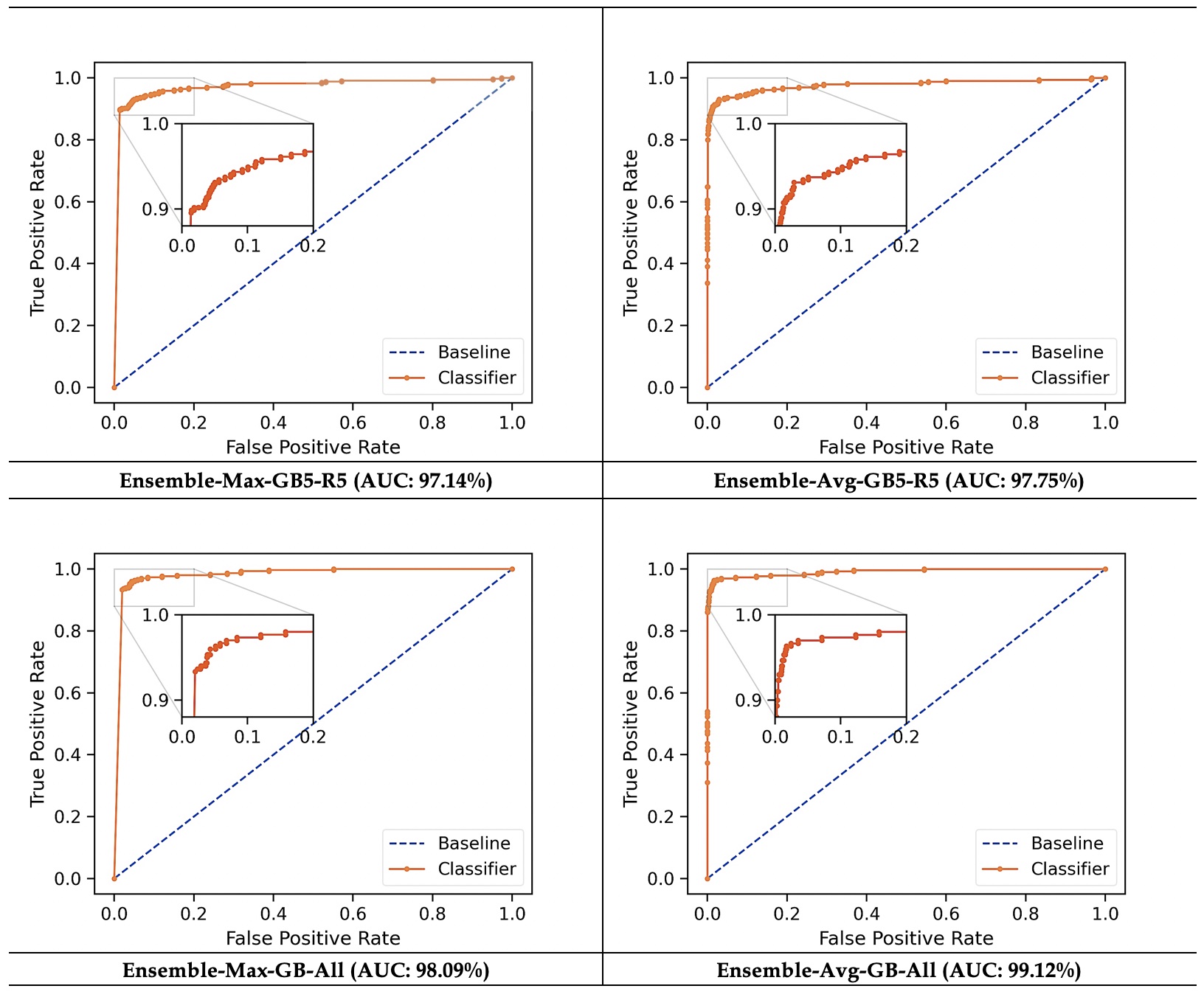}
    \caption{ROC curves for the CNN ensemble}
    \label{fig:cnn_ensemble_roc}
\end{figure}

We see that while the max-ensemble results in a greater number of false positives, it improves the recall by four points (Table \ref{tbl:cnn_ensemble}) to 94.33\% and 97.67\% for the GB5-R5 and the GB-All test set, respectively. The avg-ensemble has the advantage of having far fewer false positives and a judiciously selected prediction threshold could see false negatives further minimised for this configuration. The ROC curves for the ensemble (Figure \ref{fig:cnn_ensemble_roc}) further illustrate this trade-off. 

\begin{table}[!ht]
    \centering
    \includegraphics[width=0.48\textwidth]{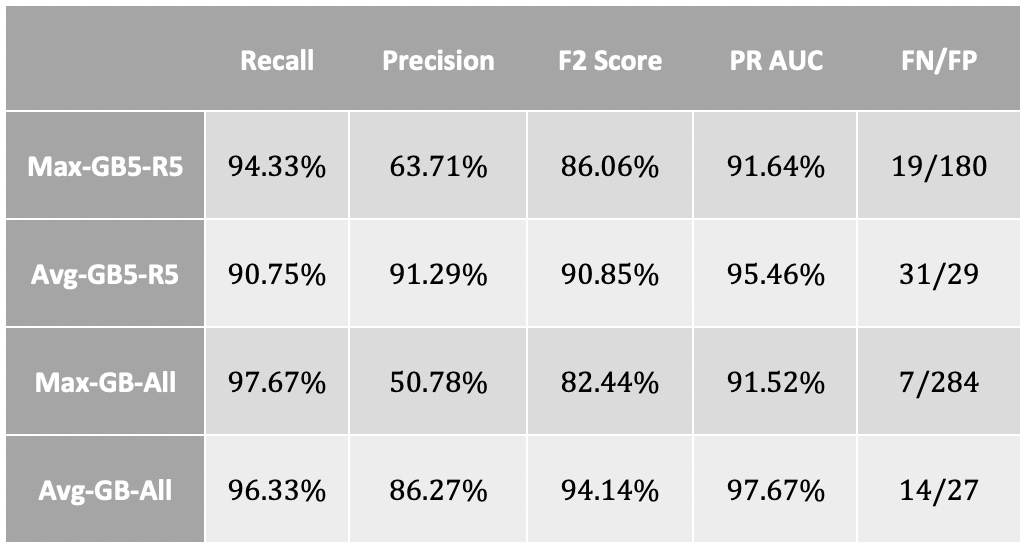}
    \caption{Evaluation metrics after configuring the five custom networks as an ensemble taken at probability threshold 0.5}
    \label{tbl:cnn_ensemble}
\end{table}

\begin{figure*}
    \centering
    \includegraphics[width=\textwidth]{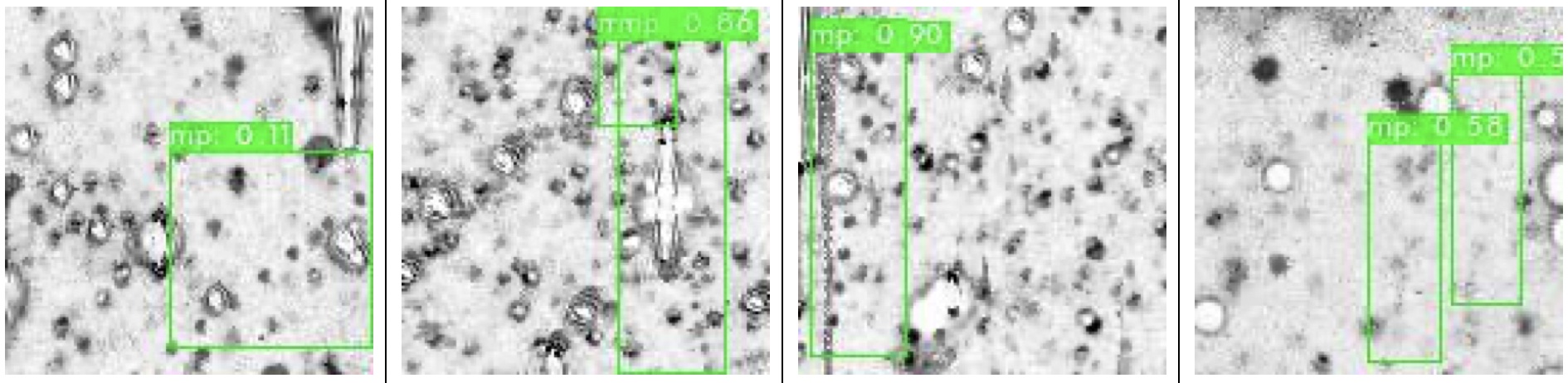}
    \caption{Tracklets localisation by YOLOv4, From left to right, these are: (538147)2016 BT90 (18.7); (67001) 1999 XN117 (19.2); (80667) 2000 BA15 (18.8); (206629) 2003 WT154 (19.0); and (281890) 2010 OA74 (20.1). Note that these images have been inverted and brightened to improve visibility (original in Figure \ref{fig:yolo_preds_apdx}).}
    \label{fig:yolo_preds_a}
\end{figure*}

The performance of object detection models is typically quantified by the mean Average Precision or mAP. The average precision (AP) measures the trade-off between precision and recall and is calculated by integrating the area that falls under the precision-recall curve for each unique values of recall where the precision value decreases. The mAP is the mean of the AP across all object classes that the model can detect, which is identical to the AP for this dataset. 
Object detection models aim to have a high overlap between the predicted and the ground truth bounding boxes (Intersection of Union or IoU) and predictions are grouped as follows:
\begin{itemize}
    \item True Positive: IoU \(>\) 0.5
    \item False Positive: IoU \(<\) 0.5 (or a duplicate) 
    \item False Negative: box not detected or the IoU \(>\) 0.5 but the object is classified wrong
\end{itemize}

The average precision is calculated as:
Calculated as:
\[AP = \Sigma(r_{(n+1)} - r_n)\tilde{p}(r_{(n+1)}) \]
\[\tilde{p}(r_{(n+1)}) = \max_{\tilde{r} \geq r_{(n+1)}}(p(\tilde{r})),\]

where \(r\) is the recall value, \(n\) represents the locations where the precision decreases, and \(\tilde{p}(r_{(n+1)})\) is the maximum precision where the \(r\) value changes. 

YOLOv4 achieved an mAP of 90.96\% with default anchor boxes and 90.95\% with custom anchor boxes on the GB5-R5 data. Neither network could detect tracklets in the GB-All data, indicating that YOLOv4 will need to be trained with labelled data from other fields and chips before it can be used to make detections in them. Some of the detections made by YOLOv4 on the GB5-R5 dataset can be seen in Figure \ref{fig:yolo_preds_a}.

The proposed model was applied towards discovering tracklets in the 543,595 images tiles from GB5-R5 that had no known asteroid tracklets. The ensemble could analyse 10,000 images in an hour and took three days to analyse all the images and proposed 50,227 candidate detections. Both version of YOLOv4 were then used for localizing tracklets in these candidate detections, with each version taking 1.5 days to make bounding box predictions. We are still analysing these candidate detections to determine if any new objects are present. 

\section{Discussion}
\label{sec:discussion}

Our classifier ensemble, trained with a relatively small amount of labelled data from just one chip in one survey field, was able to classify a set of images from unseen survey fields and chips, highlighting that it generalizes well. The success of the custom CNN architectures with fewer training parameters is perhaps down to two aspects. First, the networks might be at the optimal depth for learning to identify the pattern of fuzzy blobs that represent a tracklet. Second, the small size of the input images perhaps leads to more overfitting in the larger models. Further, rather than adding more layers, the hybrid architectures lean into adding complexity with several 1x1 convolutions. It is possible this caused the networks to learn representations that ultimately boosted the performance of the ensemble.

\begin{figure*}
    \centering
    \includegraphics[width=\textwidth]{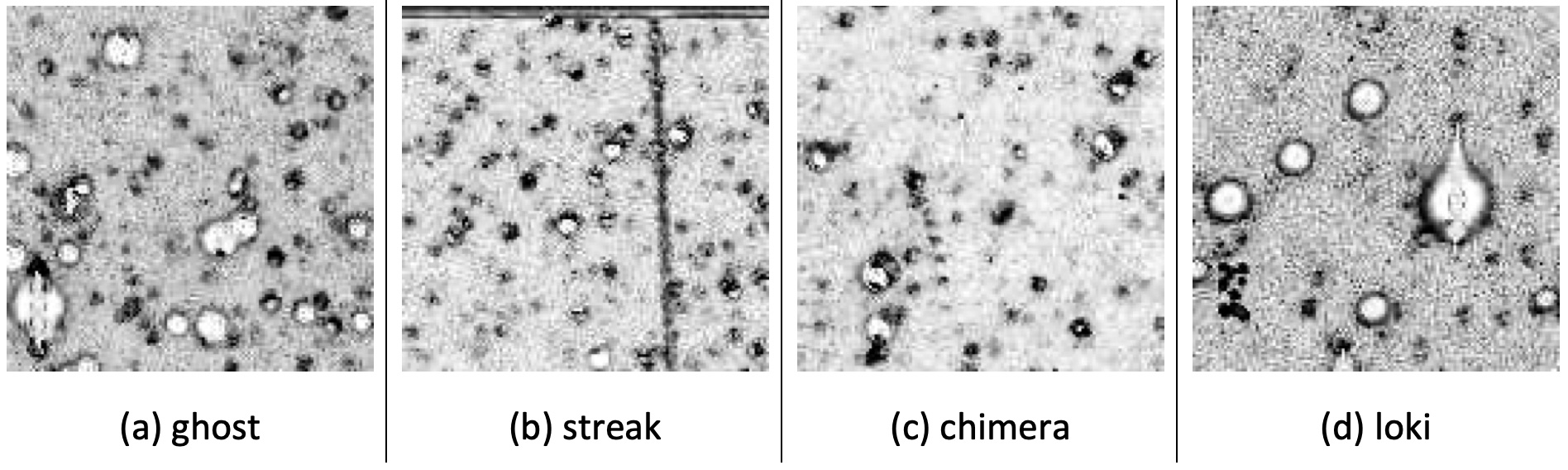}
    \caption{The false positives reported by the classification networks fall into four broad categories: (a) ghosts (short for ghosts of noise past), are the most common false positives and are a result of the additive noise from creating the composite images; (b) streaks are either satellites or other near-Earth objects or cosmic rays; (c) chimera are optical artefacts potentially caused by over-saturated stars; and (d) loki objects are artefacts that move around erratically from one observation to the next. Note that these images have been inverted and brightened to improve visibility.}
    \label{fig:cnn_fp}
\end{figure*}

\begin{figure*}
    \centering
    \includegraphics[width=\textwidth]{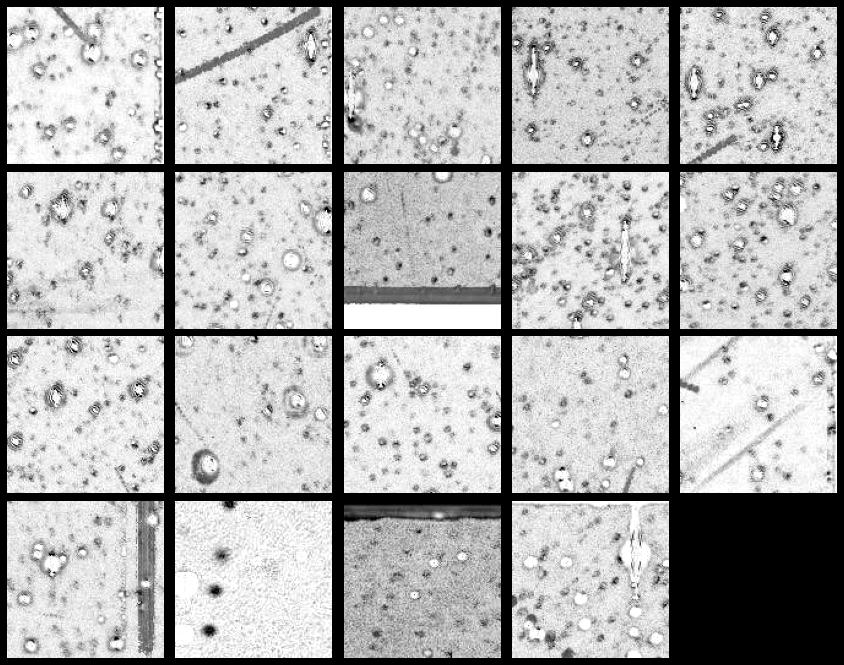}
    \caption{The false negatives reported by the classification networks are mostly either faint objects or too much like streaks. It is likely that these sorts of tracklets were not well represented in the training set. Note that these images have been inverted and brightened to improve visibility.}
    \label{fig:cnn_fn}
\end{figure*}

The false negatives (Figure \ref{fig:cnn_fn}) reported by the ensemble include instances where the tracklets resemble satellite streaks or where the point sources are slightly further apart and obscured by noise. In some cases, parts of the tracklet are on other images that have been correctly identified as a candidate detection by the ensemble. As to the false positives, they are largely cases where noise and other spurious artefacts have been mistaken for tracklets, as indicated in Figure \ref{fig:cnn_fp}. These also lead to false detections by the object detector. We believe that fine tuning the network models as more labelled data becomes available will aid in minimising misclassifications such as these.

The YOLOv4 object detector proved to be up to the task of localizing tracklets in the GB5-R5 dataset but does not generalize to unseen data. The results from the GB5-R5 set are promising and thus indicate that further training with data from the other survey fields and chips will be beneficial.

While both types of networks are currently limited by the high cadence observations they have been trained with, we are confident that, as the networks are retrained with data from other fields, they will learn to identity tracklets of slow objects or those with longer intervals between observations. Further, other surveys could use our networks, together with their pre-trained weights, as a starting point for discovering asteroids in their data. In this case, we recommend that the other surveys retrain/ fine-tune the architectures with their own data for optimal results.

\section{Conclusion}
\label{sec:conclusion}
We have shown that it is possible to train both CNN-based classifiers as well as the YOLOv4 object detector to find asteroid tracklets in the MOA difference images. The classifier ensemble proved resilient to discovering tracklets in unseen data and will be invaluable for extending the search for asteroids to the rest of the MOA-II archival data. The networks will be fine-tuned or retrained as more labelled data is available and we will investigate automating the bounding boxes required for training YOLOv4. We will also investigate working with orbit-linking software such as HelioLinC to determine the validity of the source clusters localized by YOLOv4.

While the classifiers performed well, there is potential for further improvement and development. Investigating effective denoising techniques for the stacked images or the difference images would lead to an immediate performance boost for both the classification and object detection networks. The classification network may benefit from having two inputs - perhaps the subtracted stack image along with the median stack image. This would provide the model with additional information it could use to distinguish between images with or without tracklets. Training the YOLO backbone feature extractor with the classification data first might also lead to better results.  Additionally, since much of the salient information is contained in the first layer, a small CNN based on DenseNet could also potentially be successfully trained as a classifier.

Overall, we have presented an effective toolkit for finding asteroids tracklets in the archival data of ground-based telescopes. The code for our neural network models as well as the trained weights are available at https://github.com/pcowan-astro/MOA-Asteroids. Our methodology and network architectures can be used to discover and recover asteroids in other archival survey data as well as to strengthen the analysis pipeline for current and future surveys. 

\section{Acknowledgements}
\label{sec:ack}
We thank the MOA collaboration for use of the MOA-II archival difference images. IAB acknowledges support from grant MAU1901 from the Royal Society of New Zealand - Marsden. This research has made use of data and/or services provided by the International Astronomical Union's Minor Planet Center.

\appendix
\section{Original images}
\label{sec:appendix}
\begin{figure*}
    \centering
    \includegraphics[width=.8\textwidth]{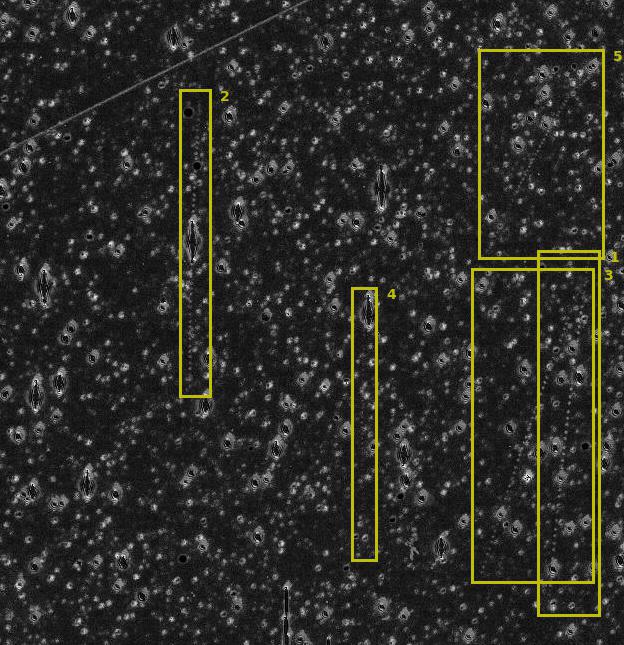}
    \caption{A stack of all 51 observations from the night of 15-May-2008 reveals 5 asteroid tracklets clearly visible in a sub-region of the stack image. Numbered in order of appearance, these are: 1 - (103842) 2000 DQ33 (19.5); 2 - (148657) 2001 SX124 (19.6); 3 - (152083) 2004 RH30 (19.9); 4 - (582743) 2016 AT221 (20.4); 5 - (338789) 2005 SZ154 (20.4). The line/streak on the top left is from a satellite.}
    \label{fig:quintet_apdx}
\end{figure*}

\begin{figure*}
    \centering
    \includegraphics[width=\textwidth]{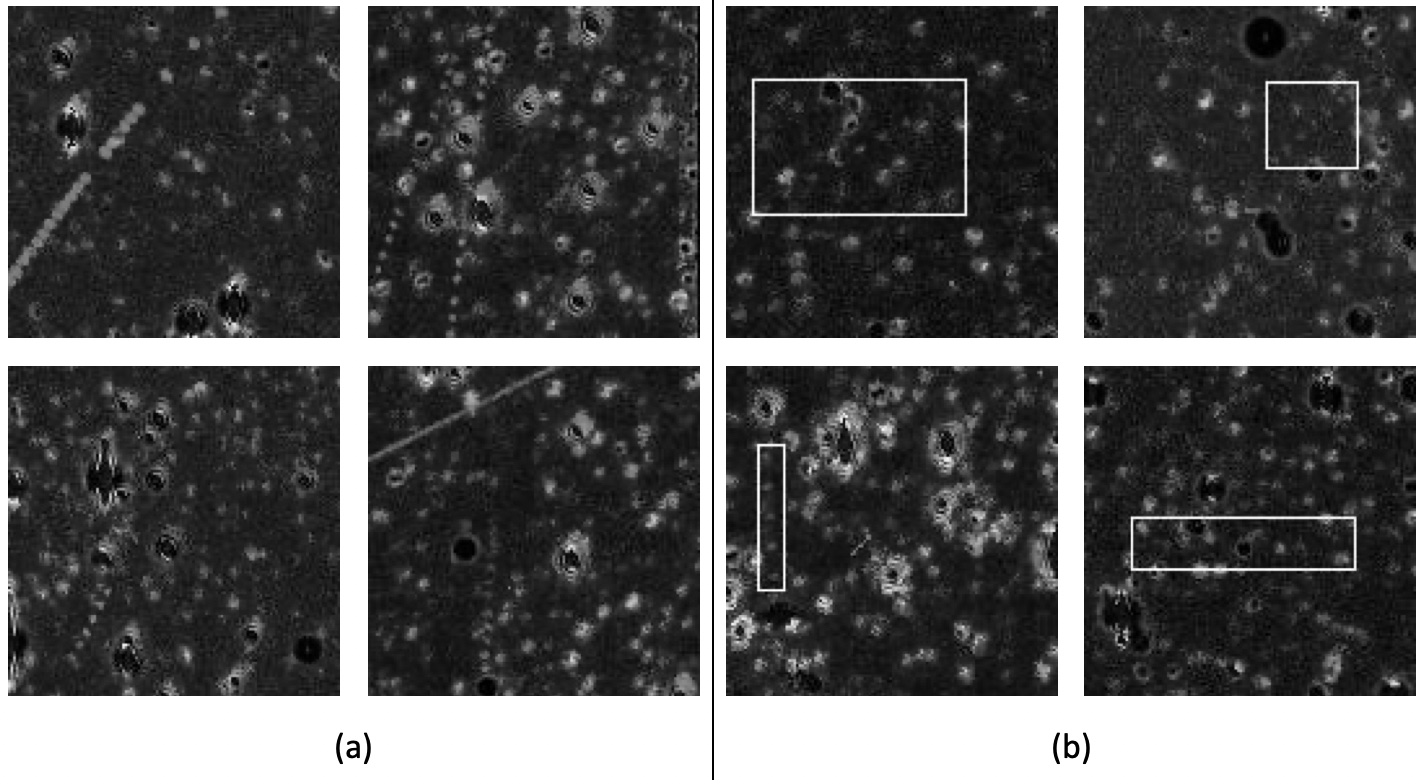}
    \caption{Asteroid tracklets seen in 128x128 sub-regions from subtracted stack images. In (b), the tracklets are localized with bounding boxes. Asteroids in the top row, left to right: (38102) 1999 JM18 (18.4); (152083) 2004 RH30 (19.9); (103842) 2000 DQ33 (19.5); (283261) 2011 FR142 (20.0); and (375674) 2009 HD5 (19.8). Asteroids in the bottom row, from left to right: (48617) 1995 HR2 (19.5); (148657) 2001 SX124 (19.6); (97948) 2000 QF124 (19.6); and (74978) 1999 TY234 (18.7).}
    \label{fig:tracklets_apdx}
\end{figure*}

\begin{figure*}
    \centering
    \includegraphics[width=\textwidth]{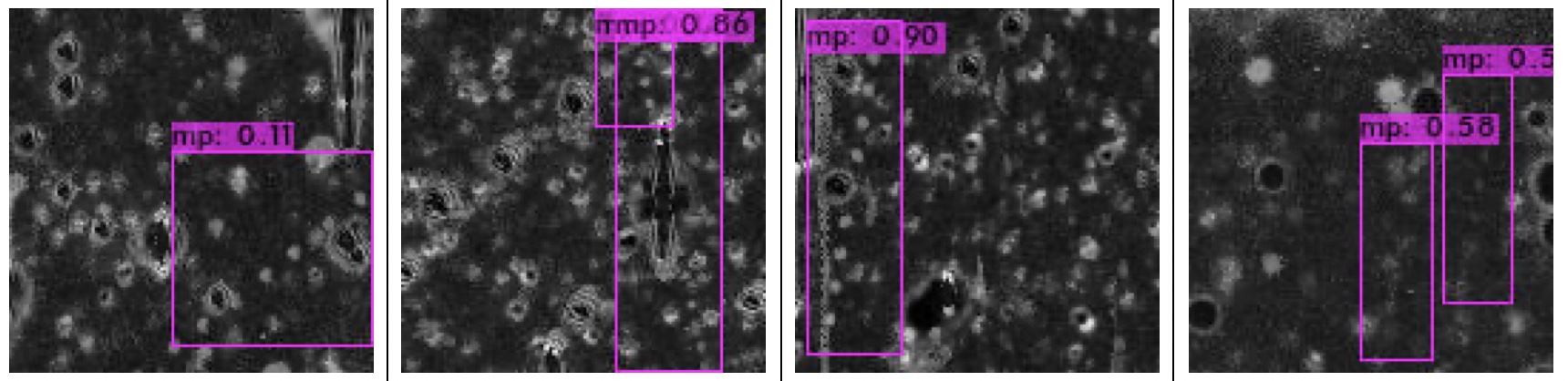}
    \caption{Tracklets localisation by YOLOv4, From left to right, these are: (538147)2016 BT90 (18.7); (67001) 1999 XN117 (19.2); (80667) 2000 BA15 (18.8); (206629) 2003 WT154 (19.0); and (281890) 2010 OA74 (20.1).}
    \label{fig:yolo_preds_apdx}
\end{figure*}

\newpage
\bibliographystyle{elsarticle-harv} 
\bibliography{references}




\end{document}